\documentstyle[12pt]{article}
\begin{document}
\begin{flushright}
MRC.PH.TH-6-96

hep-th/9604167

\end{flushright}
\newcommand{\ba}{\begin{array}}
\newcommand{\ea}{\end{array}}
\newcommand{\cR}{{\cal R}}
\newcommand{\lb}{\label}
\newcommand{\beqa}{\begin{eqnarray}}
\newcommand{\eeqa}{\end{eqnarray}}
\newcommand{\beqaa}{\begin{eqnarray*}}
\newcommand{\eeqaa}{\end{eqnarray*}}
\newcommand{\be}{\begin{equation}}
\newcommand{\ee}{\end{equation}}
\newcommand{\we}{\wedge}
\newcommand{\fr}{\frac}
\newcommand{\D}{\delta}
\newcommand{\sig}{\sigma}
\newcommand{\Sig}{\Sigma}
\newcommand{\del}{\partial}
\newcommand{\wv}{\wedge}
\newcommand{\al}{\alpha}
\newcommand{\la}{\lambda}
\newcommand{\ka}{\kappa}
\newcommand{\La}{\Lambda}
\newcommand{\ep}{\epsilon}
\newcommand{\pr}{\prime}
\newcommand{\ti}{\tilde}
\newcommand{\omo}{\omega}
\newcommand{\OO}{\Omega_{BRST}}
\newcommand{\bi}{\bibitem}
\newcommand{\cA}{{\cal A}}
\newcommand{\cL}{{\cal L}}
\newcommand{\cD}{{\cal D}}
\newcommand{\cO}{{\cal O}}
\newcommand{\cN}{{\cal N}}
\newcommand{\cS}{{\cal S}}
\newcommand{\Omo}{\Omega}

\begin{center}
{\large
CONSISTENT INTERACTIONS IN TERMS OF THE GENERALIZED FIELDS METHOD}
\vspace{3cm}

{\normalsize  \"{O}mer F. DAYI}\footnote{E-mail address: 
dayi@yunus.mam.tubitak.gov.tr} 
\end{center}
{\small \it
TUBITAK - Marmara Research Center, 
Research Institute for Basic Sciences,
Department of Physics, 
P.O. Box 21,
41470 Gebze-TURKEY }

\vspace{2cm}
\begin{center}
{\bf  Abstract}
\end{center}
{\small
The interactions which preserve the structure of 
the gauge interactions of the
free theory are introduced in terms of the generalized fields method
of solving the Batalin--Vilkovisky master equation. It is shown that
by virtue of this method the solution of the descent equations resulting 
from the cohomological analysis is provided  straightforwardly. 
The general scheme is illustrated 
by applying it to spin--1 gauge field in 3 and 4 dimensions,
to free $BF$ theory in 2--d and to the antisymmetric 
tensor field in any dimension. 
It is shown that it reproduces the results obtained by 
cohomological techniques.
}
\vspace{3cm}

\pagebreak

In a local gauge theory if the number of the physical degrees of freedom
is modified when the interactions are switched off, there may be  
some negative norm states 
which yield inconsistencies
in the scattering theory where one uses the
asymptotic states\cite{ko}.
To avoid it one deals with the local interactions which preserve
the structure of the gauge generators of the free theory.
These are the consistent 
interactions\cite{cons}. By using the cohomological aspects of the
Batalin--Vilkovisky (BV) method of quantization of 
gauge theories\cite{bv}
one obtains the descent equations whose solution yields
the consistent interactions\cite{bh}--\cite{hp}.

The generalized fields method  is proven to be powerful in 
the BV quantization of a large class of gauge 
theories\cite{gf}--\cite{or}. 
It offers the proper solution of the (BV) master equation
in a straightforward way. 

We show that when one deals with a free gauge theory
whose action can be written  first order in the exterior
derivative $d$ and bilinear in fields, the generalized fields
method can be applied.  As far as the properties of 
gauge generators are unaltered when interactions are
introduced the generalized fields method should still
be applicable. Hence, it can be utilized to obtain
the consistent interactions. Moreover, 
solution of the
related descent equations is a by-product in this scheme,
i.e. this method provides the solution of the
cohomology problem in a straightforward and compact manner. 
We do not deal with the consistent interactions 
neither non-local nor
possessing derivatives and obtainable by
field redefinitions.

The general scheme is illustrated by applying it to
the spin--1 gauge field in $d=4$ and $d=3$,
whose consistent interactions 
yield Yang-Mills and Chern-Simons field 
theories. The consistent interactions of the free $BF$ theory in
$d=2$ are shown to lead to the gauge action of the
quadratic Lie algebras\cite{ii}. 
In all of the examples the solutions of the descent 
equations are presented.
When we deal with the 
antisymmetric tensor field in D--dimension the application of the
method is shown to provide in a straightforward manner 
the solution of the cohomology problem 
obtained in \cite{hp}.

Let us deal with a free gauge theory
whose action can be written as
\be
{\cal A}_0=\int (B^a\we dA^a+\fr{1}{2}\al_{ab} B^a\we B^b)
\ee
where $\al$ can be a vanishing or  a constant matrix
which will be suppressed in the following. 
It is invariant under the infinitesimal gauge
transformations 
\be
\lb{dgi}
\D_\La A^a=d\La^a,\   \D_\La B^a =0.
\ee
To quantize this theory in terms of the BV method,  
one introduces the ghost, ghost of ghost fields
and the antifields (minimal sector),
by inspecting the reducibility properties of the 
gauge transformations. One  can attribute to each of the
fields a total form degree defined as the sum 
of the differential form degree and
the ghost number. To apply the generalized fields method 
one collects the original fields and the ones introduced for  
applying the BV method in  the generalized fields 
$\ti{A}^a,\  \ti{B}^a$ whose total form
degrees are the same with the differential 
form degrees of the original fields $A^a,\ B^a.$ 
The action written in terms
of the generalized fields
\be
S_0\equiv \int\ti{\cL}_0
 = \int  (\ti{B}^ad\ti{A}^a +
\fr{1}{2} \ti{B}^a\ti{B}^a)_{(D,0)}  \lb{S0}
\ee
(the first of the numbers in the parentheses 
is the differential form degree, the second is the ghost number
and if there is only one number it denotes the total form degree)
is the proper solution of the master equation
\be
\lb{s0s0}
(S_0,S_0) \equiv
\fr{\D_r S_0}{\D \ti{A}^a}\fr{\D_l S_0}{\del \ti{B}^a}
-\fr{\D_r S_0}{\D \ti{B}^a}\fr{\D_l S_0}{\del \ti{A}^a}=0.
\ee
To prove this observe that due to the gauge transformations
(\ref{dgi}), in $\ti{B}$ there
is not any field possessing positive ghost number
so that
$(\ti{B}\ti{B})_{(D,0)}=B\we B$ and  
there is not $B^\star$ term in 
$(\ti{B}d\ti{A})_{(D,0)}.$ 

BRST transformation of a functional is defined as
\be
\Omo_0F(\ti{A},\ti{B})=\fr{\del_rF}{\del \ti{A^a}}\Omo_0\ti{A}^a
+\fr{\del_rF}{\del \ti{B^a}}\Omo_0\ti{B}^a
\ee
where 
\be
\lb{brstf}
\Omo_0\ti{A}^a=-\fr{\del_lS_0}{\del \ti{B}^a},\  
\Omo_0\ti{B}^a=\fr{\del_lS_0}{\del \ti{A}^a}.
\ee

In terms of the BRST charge $\Omo_0$
(\ref{s0s0})  can equivalently be formulated as
\be
\lb{nls}
\Omo_0\ti{\cL}_0 +d\omo_{(D-1,1)}=0,
\ee
where $\omo_{(D-1,1)}$ is a suitable $(D-1)$--form possessing ghost number
one. Moreover, due to the fact that $(\Omo_0 + d)^2=0,$
there are the descent equations 
\be
\lb{nd}
\Omo_0 \omo_{(D-1,1)} +d\ka_{(D-2,2)}=0,\ 
\Omo_0 \ka_{(D-2,2)} +d\la_{(D-3,3)}=0,\ 
\cdots,
\ee
with suitable $\ka ,\la ,\cdots,$ vanishing at a certain step.

Obviously $\Omo_0 $ maps $\ti{A}\rightarrow \ti{dA}$ and
 $\ti{B}\rightarrow \ti{dB}$ except 
\[
\Omo_0 B^{\star a} =dA^a +B^a.
\]
However, there is not any $B^\star$ term in $\ti{\cL}_0 ,$ so that, 
the BRST transformation of the free 
(\ref{nls}) theory can be generalized as
\be
\lb{des1}
(\Omo_0 +d) (\ti{B}^ad\ti{A}^a)_{(D)}=0.
\ee
We dropped the $B^a\we B^a$ term which does not transform under $\Omo_0.$
(\ref{des1}) written in components produces (\ref{nd}) with the 
identification $\omo_{(D-1,1)}\equiv (\ti{B}^ad\ti{A}^a)_{(D-1,1)},$
$\ka_{(D-2,2)}\equiv (\ti{B}^ad\ti{A}^a)_{(D-2,2)}$ and so on.
A compact notation similar to (\ref{des1}) is also presented
in \cite{dr}

The consistent interactions will be introduced as cubic or higher
order in the generalized fields $\ti{\Phi} \equiv (\ti{A},\ti{B})$:
\beqa
S_{int} & \equiv & \sum_{k=1,M=3}g^k_MS_{Mk} =\int 
\sum_{M=3}(g_M{\ti{\Phi}}^M+g_M^2{\ti{\Phi}}^{M+1} 
+\cdots )_{(D,0)} \nonumber \\
& = & (g_3{\ti{\Phi}}^3+g_3^2{\ti{\Phi}}^4 +\cdots 
+g_4\ti{\Phi}^4+g_4^2\ti{\Phi}^5+\cdots )_{(D,0)},
\eeqa
such that 
\be
\lb{me}
(S_0+S_{int} , S_0 +S_{int} )=0.
\ee
$g_M$ are independent coupling constants, hence,
(\ref{me}) leads to 
\beqa
(S_0 , S_0 ) & = & 0, \\
(S_0 ,S_{M1}) & = & 0, \lb{yp1} \\
2 (S_0,S_{M2})+(S_{M1}, S_{M1})& = &0,  \\
. & & \nonumber \\
. & & \nonumber \\
. & & \nonumber
\eeqa
(\ref{yp1}) can equivalently 
be written in terms of the BRST charge $\Omo_0$ as
\be
\lb{ns}
\Omo_0 {\ti{\Phi}}^M_{(D,0)}
+d k_{(D-1,1)} =  0,
\ee
where $k$ is a suitable $(D-1)$--form possessing ghost number one.
As it has already mentioned
$\Omo_0 $ maps $\ti{A}\rightarrow \ti{dA}$ and
$\ti{B}\rightarrow \ti{dB}$ except $B^\star .$ Let 
the $B^\star $ dependent terms in $S_{M1}$ are
\[
({\ti{\Phi}}^M_{(D,0)})_{B^\star} \sim 
(B^\star{\ti{\Phi}}^{M-1} +B^{\star2} \ti{\Phi}^{M-2}
+\cdots +B^{\star M-1}{\ti{\Phi}})_{(D,0)}.
\]
Thus in (\ref{ns}) the terms which does not possess an exterior derivative 
will be
\be
\lb{st}
\Omo_0{\ti{\Phi}}^M_{(D,0)}+dk_{D-1,1} 
\sim B{\ti{\Phi}}^{M-1} +2B B^\star \ti{\Phi}^{M-2}
+\cdots +(M-1)B^{\star M-2}\ti{\Phi}.
\ee
There are two possibilities 
$i)$ (\ref{st}) vanishes due to symmetry properties, so that 
(\ref{ns}) is satisfied with $k_{(D-1,1)}\equiv 
{\ti{\Phi}}^3_{(D-1,1)}, $
$ii)$  (\ref{st}) does not vanish so that, (\ref{ns}) is not 
satisfied. In the latter case the interactions are not consistent.
Therefore, we conclude that as far as we deal with
the consistent interactions (\ref{yp1}) is equivalent to 
\be
\Omo_0 {\ti{\Phi}}^M_{(D,0)}
+d {\ti{\Phi}}^M_{(D-1,1)}  =  0,\lb{ms1} 
\ee
and it can be generalized to $D$--total form:
\be
(\Omo_0 +d) {\ti{\Phi}}^M_{(D)} =  0,
\ee
which leads to the descent equations
\beqa
\Omo_0 {\ti{\Phi}}^M_{(D,0)}
+d {\ti{\Phi}}^M_{(D-1,1)} & = & 0,\lb{d1} \\
\Omo_0 {\ti{\Phi}}^M_{(D-1,1)}
+d {\ti{\Phi}}^3_{(D-2,2)} & = & 0,\lb{d2} \\
. && \nonumber \\
. && \nonumber \\
\Omo_0 {\ti{\Phi}}^M_{(1,D-1)}
+d {\ti{\Phi}}^M_{(0,D)} & = & 0,\lb{dD-1} \\
\Omo_0 {\ti{\Phi}}^M_{(0,D)}
& = & 0.\lb{dD} 
\eeqa
Obviously some of $\ti{\Phi}^{M}_{(a,b)}$ can be vanishing.
We conclude that the generalized fields method provide the 
consistent interactions as well as the solution of the descent
equations.

Gauge invariance of the interacting theory can be obtained by replacing 
$S_0$ with $S_0+S_{int}$ in (\ref{brstf}) on the surface where the ghost 
fields and the antifields are vanishing.

\noindent
\underline{{\it Spin--1 Gauge Field in D=4:   }}
Free theory given by the first order action
\be
\lb{eym}
\cL_0
=\fr{-1}{2}\int d^4x\   ( B^{a\mu \nu}   
(\del_\mu A_\nu^a - \del_\nu A_\mu^a )
-\fr{1}{2} B_{\mu \nu}^a B^{a\mu \nu} ),
\ee
is invariant under the 
infinitesimal gauge transformations
\[
\D_\La A^a_\mu =\del_\mu \La^a\ ,\  \D_\La B^a_{\mu \nu}= 0.
\]
The theory is irreducible, so
that  we 
need to introduce (in the minimal sector) the ghost field
$\eta_{(0,1)}$, and the antifields $A^\star_{(3,-1)},$
$\eta^\star_{(4,-2)},$ and $B^\star_{(2,-1)}.$ 
Here the star indicates
the antifields as well as the Hodge-map. 

The generalized fields are
\[
\begin{array}{lcl}
\ti{A}^a & =  &  A_{(1,0)}^a+\eta^a_{(0,1)} + B_{(2,-1)}^{\star a} , \\
\ti{B}^a & = & - A^{\star a}_{(3,-1)}- \eta^{\star a}_{(4,-2)}+B_{(2,0)}^a.
\end{array}
\]

The proper solution of the master equation is
\beqa
\lb{obt}
S_0 & = & \fr{-1}{2}\int d^4x\   
(\ti{B}^a d\ti{A}^a -\fr{1}{2} \ti{B}^a\ti{B}^a 
)_{(4.0)}, \\
& = & -\int d^4x\ (\fr{1}{2}B^{a\mu \nu}
(\del_\mu A^a_\nu -\del_\nu A^a_\mu ) -
A_\mu^{\star a} \del^\mu \eta^a 
-\fr{1}{4} B_{\mu\nu}^a B^{a \mu\nu}), \lb{syme}
\eeqa
from which we read the BRST transformations
\be
\begin{array}{lll}
\Omo_0A^a_\mu =\del_\mu \eta^a; &   \Omo_0 B^{a\mu\nu}=0; &
\Omo_0 \eta^a =0;\\ 
\Omo_0 A^{\star a\mu} =\del_\nu B^{a\mu \nu}; &
\Omo_0 B^{\star a}_{\mu \nu} =
\del_\mu A^a_\nu -\del_\nu A^a_\mu  -B^a_{\mu \nu}; &
\Omo_0\eta^{a \star}=-\del_\mu A^{\star a \mu}.  
\end{array}
\ee

The unique candidate for a consistent cubic interaction is
the 4--total form 
\be
\lb{cin}
\ka_{(4)} \equiv \fr{1}{2}f_{abc}\ti{B}^a\ti{A}^b\ti{A}^c.
\ee
One can show that, in fact, (\ref{cin}) provides the solution 
of the descent equations
\beqa
\Omo_0\kappa_{(4,0)} +
d \kappa_{(3,1)}=0, \nonumber \\
\Omo_0\kappa_{(3,1)} +
d \kappa_{(2,2)}=0, \nonumber \\
\Omo_0\kappa_{(2,2)} =0, \nonumber \\
\kappa_{(1,3)} =
 \kappa_{(0,4)}
=0, \nonumber 
\eeqa
if $f_{abc}$ is totally antisymmetric. Written explicitly
\beqaa
\kappa_{(4,0)} & = & f_{abc} 
(\fr{1}{2}B^{a\mu\nu} A^b_\mu A^c_\nu 
- A^{\star a\mu} A_\mu^b \eta^c
+ B^{a\mu\nu}\eta^b B^{\star c}_{\mu\nu}  
- \fr{1}{2}\eta^{\star a} \eta^b \eta^c ), \\ 
\kappa_{(3,1)} & = & f_{abc} 
(B^{\mu\nu}A_\nu^b\eta^c +\fr{1}{2}A^{\star a\mu}\eta^b \eta^c ), \\
\kappa_{(2,2)} & = &-\fr{1}{2} f_{abc}B^{a\mu\nu}\eta^b\eta^c.
\eeqaa
If $f_{abc}$  satisfy the Jacobi identities,
one can show that $S_{31}=g_3\int \ka_{(4,0)}$ satisfies
\be
(S_{31},S_{31})=0.
\ee
Hence, there is no need of adding a quartic interaction with $g_3^2$
coupling constant. However, one can in principle add
a 4--total form $\ti{A}^4$ with another coupling constant
$g_4.$ For being a consistent interaction
$S_{41} $ should be BRST invariant. 
But one can show that
\[
\Omo_0 \ti{A}^4_{(4,0)} \neq dK_{(3,1)},
\]
for any $K.$ Therefore, we cannot add this interaction term.
Observe that after a gauge fixing $B^\star =0$
and using the equations of motion related to $B^{\mu \nu},$
Yang-Mills theory follows with the required quartic interaction.

\noindent
\underline{ {\it Spin--1 Gauge Field in D=3:} }
The free gauge theory is
\be
\cA_0=\int  A^a\wedge dA^a.
\ee
It is invariant under the infinitesimal
gauge transformations $\delta_\La A^a =d\La^a .$
It is an irreducible theory,
so that the generalized field is
\be
\ti{A}^a=A_{\mu (1,0)}^a
+\eta^a_{(0,1)} +A^{\star a}_{\mu (2,-1)}
+\eta^{\star a}_{(3,-2)}.
\ee
The solution of the master equation 
\be
\lb{css}
S_0 =\int \ti{A}^a d\ti{A}^a,
\ee
leads to the BRST transformations
\beqaa
\Omo_0 A_\mu^a=\del_\mu \eta^a ,& \Omo_0\eta^a=0 , \\
\Omo_0A^{\star a \mu \nu}=-\ep^{\mu \nu \rho}\del_\nu A^a_\rho, &
\Omo_0\eta^{\star a}= -\del_\mu A^{\star a \mu}.
\eeqaa

The unique candidate for a consistent interaction is 
the ghost zero component of the 3--total form
$\xi\equiv \ti{A}^3.$
Indeed, one can show that 
if $f_{abc}$ are antisymmetric in all of the indices 
the descent equations
\be
(\Omo_0+d)(f_{abc}\ti{A}^a \ti{A}^b \ti{A}^c)_{(3)}=0,
\ee
are satisfied, where the components are
\beqaa
\xi_{(3,0)} & = &
-f_{abc}(A^a \wedge A^b \wedge A^c 
+6A^{\star a} \wedge A^b \wedge \eta^c  
+3\eta^{\star a}\eta^b\eta^c ) ,\\
\xi_{(2,1)} & = &
3f_{abc}(A^a \wedge A^b \wedge \eta^c 
+A^{\star a} \wedge \eta^b \wedge A^c ), \\
\xi_{(1,2)} & = &
3f_{abc}A^a \wedge \eta^b \wedge \eta^c      , \\
\xi_{(0,3)} & = &
-f_{abc}\eta^a \wedge \eta^b \wedge \eta^c .
\eeqaa

There is no need of any further interaction term because, 
the total action $S =S_0 +g_3\int \xi_{(3,0)}$ satisfies
the master equation if $f_{abc}$ satisfy the Jacobi identities.

\noindent
\underline{{\it Free $BF$ Theory in D=2:} }
Deal with the free gauge
theory given by the Lagrange density
\be
\lb{l0}
{\cal L}_0=-\fr{1}{2}\ep^{\mu \nu}\Phi^a(\del_\mu h_\nu^a
-\del_\nu h_\mu^a),
\ee
which is invariant under the 
infinitesimal gauge transformations
\be
\delta h_\mu^a  = \del_\mu \la^a ,\
\delta \Phi^a  = 0 .
\ee
Although a careful treatment of the global modes showed 
that at some points of the target manifold the theory
is reducible\cite{ss}, we deal with the regions of the
target manifold which does not include these points. Hence,
the generalized fields are
\beqa
\ti{h}^a 
= h^a_{(1,0)} +\eta^a_{(0,1)}+\Phi^{\star a}_{(2,-1)}, \\
\ti{\Phi}^a
=-h^{\star a}_{(1,-1)}-\eta^{\star a}_{(2,-2)}+\Phi^a_{(0,0)}.
\eeqa
The BV quantized free action
\be
S_0=\fr{1}{2}\int \ti{\Phi}^ad\ti{h}^a ,
\ee
leads to the BRST transformations
\beqaa
\Omo_0 h_\mu^a=\del_\mu\eta^a, &
\Omo_0 \Phi^a=0, &
\Omo_0  \eta^a =0, \\
\Omo_0 h^{\star a\mu}=\ep^{\mu\nu}\del_\nu\Phi^a, &
\Omo_0 \Phi^{\star a}=\ep^{\mu\nu}\del_\mu h_\nu^a, &
\Omo_0  \eta^{\star a} =-\del_\mu h^{\star a \mu }.
\eeqaa
There is a unique candidate for a consistent cubic interaction:
\be
\lb{qlas}
S_{31}=g_3\int \sig_{(2,0)}\equiv 
g_3\fr{1}{2}\int d^2x f_{abc}
(\ti{\Phi}^a\ti{h}^b\ti{h}^c)_{(2,0)} .
\ee
In fact, one can show that $\sig_{(D)}$ written in components
\beqaa
\sigma_{(2,0)} & = &
f_{abc}(
\fr{1}{2}\ep^{\mu\nu}h_\mu^ah_\nu^b\Phi^c
-h^{\star a \mu}h_\mu^b\eta^c
+\Phi^{\star a}\Phi^b \eta^c
-\fr{1}{2}\eta^{\star a} \eta^b \eta^c ), \\
\sigma_{(1,1)} & = &
f_{abc}( h^a_\mu \eta^b\Phi^c -\fr{1}{2}h^{\star a}_\mu \eta^b \eta^c), \\
\sigma_{(0,2)} & = &
-\fr{1}{2}f_{abc} \eta^a\eta^b\Phi^c ,
\eeqaa
satisfy the descent equations if $f_{abc}$
is antisymmetric in all of the indices. Moreover,
if they satisfy the Jacobi identities one can show that
\[
(S_1,S_1)=0.
\]
Although, there is no need of quartic interaction with a coupling
constant $g_3^2,$ one can in principle add  
quartic or higher interaction terms
\[
\sum_{M=4} g_MS_{M1} \equiv \sum_{M=4} g_M 
\int (\ti{\Phi}^{M-2}\ti{h}^2)_{(2,0)},
\]
if they satisfy the descent 
equations (\ref{d1})--(\ref{dD}). Let us deal with the quartic term
\be
S_{41}=  g_4\int d^2x 
V_{bc}^{ad}(\ti{\Phi}_a 
\ti{\Phi}_d\ti{h}^b\ti{h}^c)_{(2,0)}.
\ee
Indeed one can show that
\beqa
(\Omo_0 +d)
(V_{bc}^{ad} \ti{\Phi}_a 
\ti{\Phi}_d\ti{h}^b\ti{h}^c)_{(2)}& = & 0, \\
(S_1,S_2) & = & 0, \\
(S_2,S_2) & = & 0,
\eeqa
if the constants satisfy
\beqa
\lb{sym}
V_{ab}^{cd}=-V_{ba}^{cd},\  
V_{ab}^{cd}=V_{ab}^{dc}, \\
{f_{[ab}}^dV_{c]d}^{ef}  +V_{[ab}^{df}{f_{c]d}}^e + 
V_{[ab}^{ed}{f_{c]d}}^f =0,   \\
V_{[ab}^{de}V_{c]d}^{fg}  =0,   \lb{ji} 
\eeqa
where $[\ ]$ denotes that the indices within them are 
antisymmetrized.

The higher interactions can also be treated similarly.

\noindent
\underline{{\it Antisymmetric Tensor Field in D--dimension:} }
The first order free action of the two form field
$B=B_{\mu\nu}dx^\mu\we dx^\nu$ in any D dimensions is
\be
\lb{ast}
\cA_0=\int (dB^a\we H^a -\fr{1}{2}\hat{H}^a\we H^a),
\ee
where $H$ is a $D-3$ differential form and $\hat{H}$ 
is its Hodge dual. (\ref{ast}) is invariant under the
gauge transformations
\be
\lb{rgi}
\D_\La B^a =d \La^a ,
\ee
where $\La^a$ is 1-form. The gauge transformations (\ref{rgi})
are reducible because, they vanish for
$\La^a =d\ep^a ,$ where $\ep^a$ is a scalar. Thus, in the minimal sector
there are ghost and ghost of ghost fields. The 
generalized fields are
\beqaa
\ti{B}^a_{(2)} & = & B^a_{(2,0)}
+C^a_{(1,1)}+\eta^a_{(0,2)}-H^{\star a}_{(3,-1)}, \\
\ti{H}^a_{(D-3)} & = & B^{\star a}_{(D-2,-1)}+C^{\star a}_{(D-1,-2)}
+\eta^{\star a}_{(D,-3)}+H^a_{(D-3,0)} ,
\eeqaa
where, now, star indicates also the Hodge map.

Solution of the master equation for the free theory is
\be
\lb{sl0}
S_0=\int (d\ti{B}^a\ti{H}^a -\fr{1}{2}\hat{H}^a \we H^a)_{(D,0)}.
\ee
Hence, the BRST transformations are 
\[
\ba{llll}
\Omo_0B^a=dC^a, &\Omo_0H^a= 0, &\Omo_0C^a= d\eta^a ,
& \Omo_0\eta^a= 0,\\
\Omo_0B^{\star a}=dH, &\Omo_0H^{\star a}=dB -\hat{H} ,
&\Omo_0C^{\star a} = -dB^{\star a},
&\Omo_0 \eta^{\star a}= dC^{\star a}.
\ea
\]

There are four combinations of the generalized fields which 
are cubic: 

\noindent
1) $\ti{H}^3$ is a $3(D-3)$--total form. For being 
a Lagrange density the form degree should be $D,$ 
which means that the dimension should be
$D=9/2.$ Hence, it is not permitted.

\noindent
2) $\ti{B}^3$ is 6--total form so it is permitted in $D=6.$
In 6--dimension 
\[
\ti{B}^3_{(6,0)} =f_{abc}(B^a\we B^b \we B^c +6B^a \we C^b \we H^{\star c}
+2\eta^{\star a}\we H^{\star b} \we H^{\star c}).
\]
However, one can easily observe that
\[
\Omo_0\ti{B}^3_{(6,0)}\neq d\ka_{(5,1)}
\]
for any $\ka_{(5,1)}.$ Thus this is excluded.

\noindent
3) $\ti{H}\ti{B}^2$ is a (D+1)--total form. 
Therefore it cannot be a Lagrange density.

\noindent
4) $\ti{H}^2 \ti{B}$ is $(2D-4)$--total form thus
it is only permitted in $D=4.$ 

Before dealing with the case 4
let us see if there can be some other consistent
interactions which are quartic or higher.
There can be
\[
(\ti{H}^m \ti{B}^n)_{(D,0)};\,\,  m,n\leq 1,\  m+n>3,
\]
which is a $(m(D-3)+2n)$--total form. Thus the dimensions should be
\[
D=\fr{3m-2n}{m-1},
\]
if the interactions are consistent.
However, one can easily observe that there is not any $n,\ m$
which lead to an acceptable dimension. Hence we can conclude that
the unique candidate for a consistent interaction is the 
case 4.

In fact in $D=4$, $\omo_{(4)}\equiv \ti{H}^2\ti{B}$ leads to
the descent equations
\beqaa
\Omo_0\omo_{(4,0)} +d\omo_{(3,1)} & = & 0, \\
\Omo_0\omo_{(3,1)} +d\omo_{(2,2)} & = & 0 ,\\
\Omo_0\omo_{(2,2)} & = & 0, \\
\omo_{(1,3)}  = \omo_{(0,4)} & = & 0.
\eeqaa
The components written explicitly are
\beqaa
\omo_{(4,0)} & = &f_{abc}(H^a\we H^b\we B^c 
+2H^a\we B^{\star b} \we C^a
-2H^a\we C^{\star b}\we \eta^c 
+B^{\star a}\we B^{\star b}\we \eta^c), \\
\omo_{(3,1)} & = &f_{abc}(-H^a \we H^b\we C^c
+2H^a\we B^{\star b}\we \eta^c) \\
\omo_{(2,2)} & = &f_{abc}H^a \we H^b \we \eta^c,
\eeqaa
where $f_{abc}$ is totally antisymmetric in its indices.
Moreover, one can show that if $f_{abc}$ satisfy the 
Jacobi identities 
\[
S_{31} =g_3\int \omo_{(4,0)},
\]
satisfies the master equation:
\[
(S_{31},S_{31})=0.
\]
The result which we obtained agree with \cite{hp},
where the same problem is considered in terms of cohomological
techniques.

\pagebreak

\newcommand{\bpl}{ Phys. Lett. }
\newcommand{\mpl}{ Mod. Phys. Lett. }
\newcommand{\np}{ Nucl. Phys. }
\newcommand{\phr}{ Phys. Rep. }
\newcommand{\dpre}{ Phys. Rev. }

\end{document}